\documentclass[11pt]{article}
\usepackage{blois}

\def\be{\begin{equation}}
\def\ee{\end{equation}}
\def\bea{\begin{eqnarray}}
\def\eea{\end{eqnarray}}

\usepackage{amsmath,amsfonts}
\long\def\symbolfootnote[#1]#2{\begingroup
\def\thefootnote{\fnsymbol{footnote}}\footnote[#1]{#2}\endgroup}

\newcommand{\Photo}{}

\begin{document}
\begin{flushright}
MITP /13-002
\end{flushright}
\vspace*{4cm}
\title{Measuring the Higgs boson self-coupling at the LHC using ratios of cross sections}

\author{FLORIAN GOERTZ$^{b\hspace{0.5mm}}$, ANDREAS PAPAEFSTATHIOU$^{c\hspace{0.5mm}}$, LI LIN YANG$^{a\hspace{0.5mm}}$ and JOS\'E FRANCISCO ZURITA$^{d\hspace{0.5mm}}$\symbolfootnote[1]{speaker}}
\address{$^a$ School of Physics and State Key Laboratory of Nuclear Physics and Technology, Peking University, Beijing 100871, China.\\
$^b$ Institut f\"ur Theoretische Physik, ETH Z\"urich, 8093 Zurich,
Switzerland. \\
$^c$  Institut f\"ur Theoretische Physik, Universit\"at Z\"urich,
8057 Zurich, Switzerland.\\
$^d$ PRISMA Cluster of Excellence \& Mainz Institute for Theoretical Physics Johannes Gutenberg University, 55099 Mainz, Germany.
}

\maketitle\abstracts{We consider the ratio between the double and single Higgs production cross sections and examine the prospect of measuring the trilinear Higgs self-coupling using this observable. Such a ratio has a reduced theoretical (scale) uncertainty than the double Higgs cross section. We find that with 600 fb$^{-1}$, the 14 TeV LHC can constraint the trilinear Higgs self coupling to be positive, and with 3000/fb one could measure it with a ${+30  \%} ({-20  \%})$ accuracy.}

After electroweak symmetry breaking the Higgs potential is a quartic polynomial in $H$, 
\be
V(H) = \frac{1}{2} M_H^2 H^2 + \lambda_{HHH} v H^3 + \frac{1}{4} \lambda_{HHHH} H^4 \, ,
\ee
where $H$ is the Higgs field. In the Standard Model (SM) $\lambda_{HHH}^{\rm SM} = \lambda_{HHHH}^{\rm SM} =  m_H^2 / 2 v^2 \sim 0.13 $.
The first coefficient (essentially the Higgs mass) was determined by discovering the Higgs~\cite{CMS_Higgs,ATLAS_Higgs}. To extract the trilinear and quartic coefficients one needs double and triple Higgs production, respectively.
The latter is very difficult to measure at the LHC, due to its very small cross section~\cite{Plehn:2005nk}. But the trilinear Higgs coupling might be extracted with some accuracy by the 14 TeV LHC, and our aim is to study how precisely it can be done~\cite{Goertz:2013kp}. 

\begin{figure}[!htp]
\begin{minipage}{0.32\linewidth}
\centerline{\includegraphics[width=0.9\linewidth]{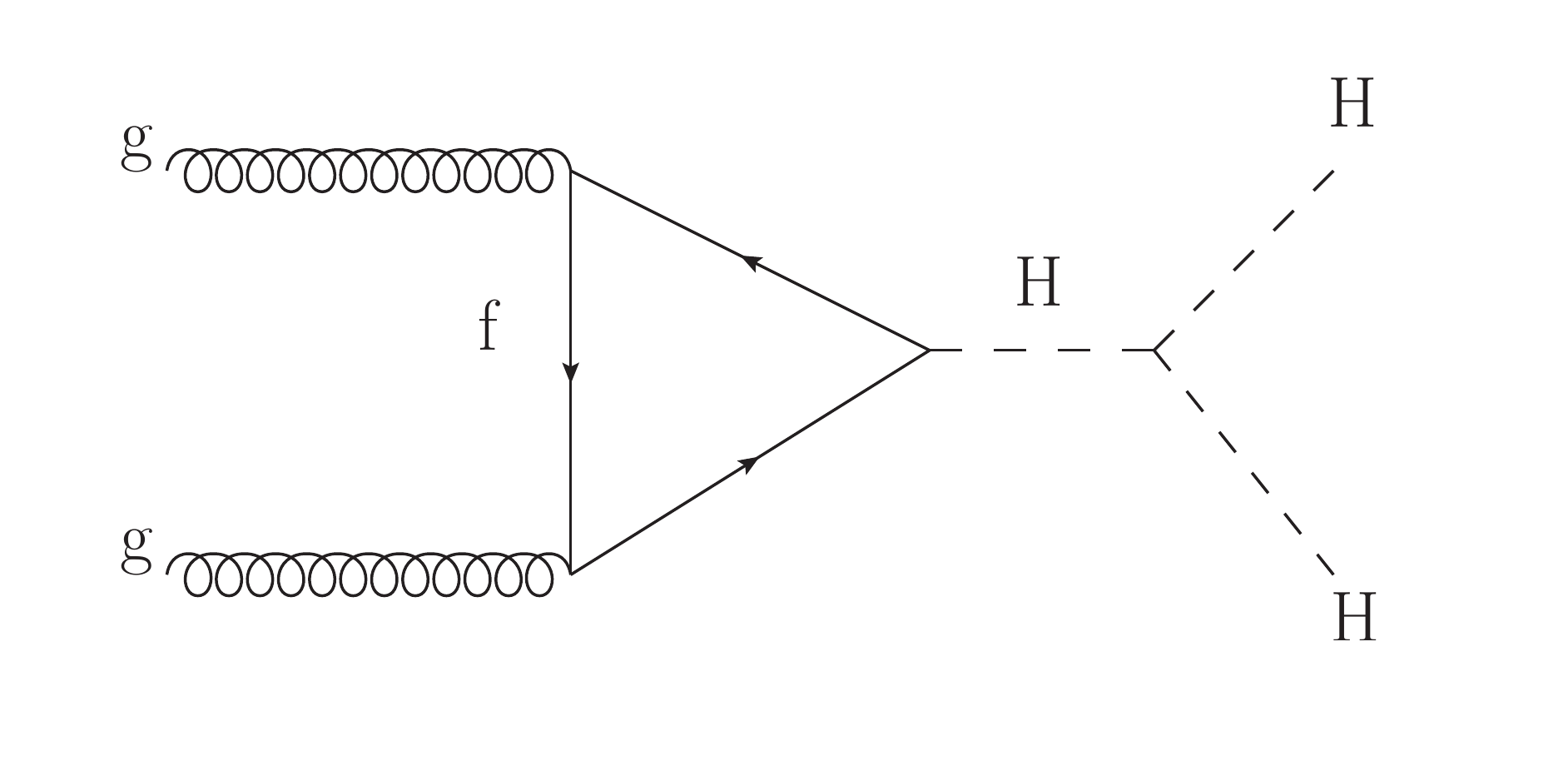}}
\end{minipage}
\hfill
\begin{minipage}{0.32\linewidth}
\centerline{\includegraphics[width=0.9\linewidth]{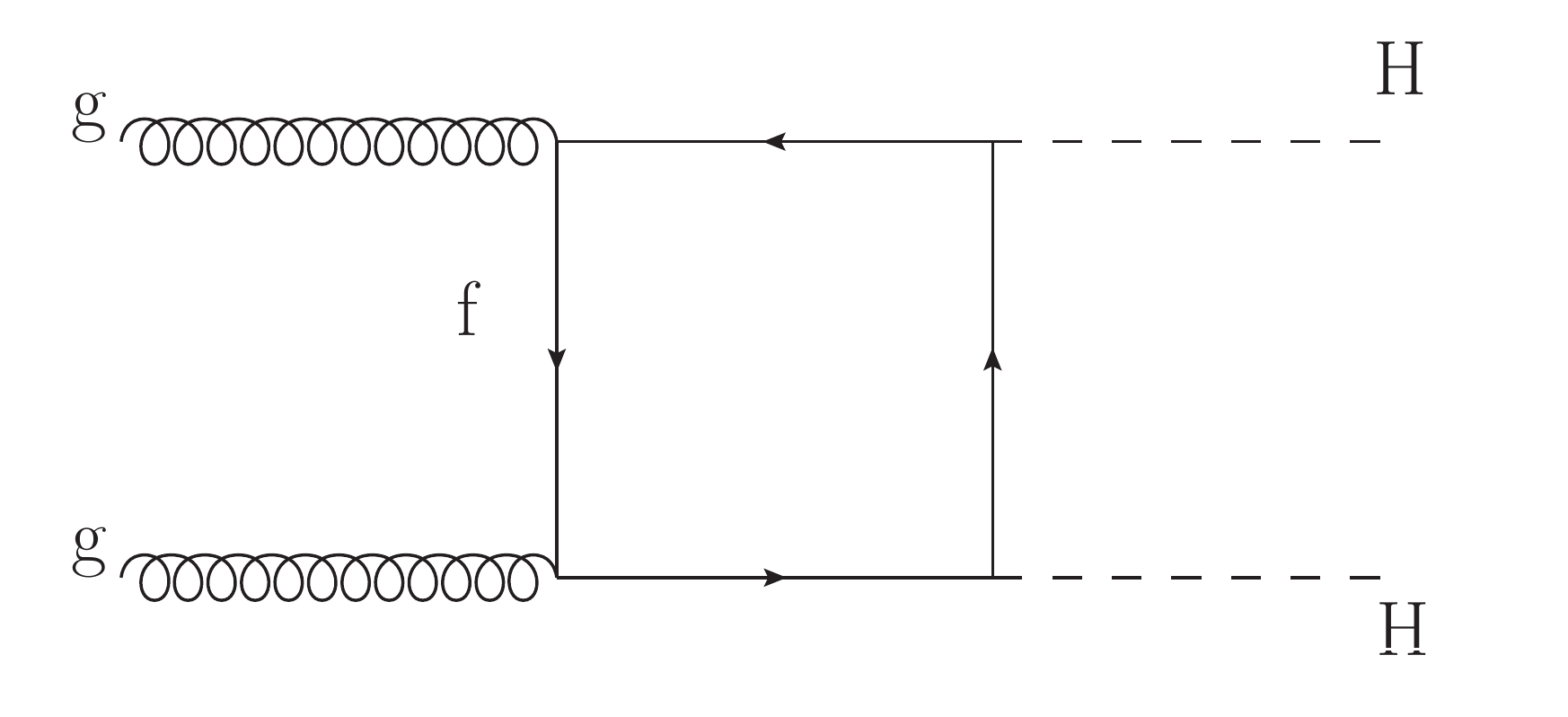}}
\end{minipage}
\hfill
\begin{minipage}{0.32\linewidth}
\centerline{\includegraphics[width=0.9\linewidth]{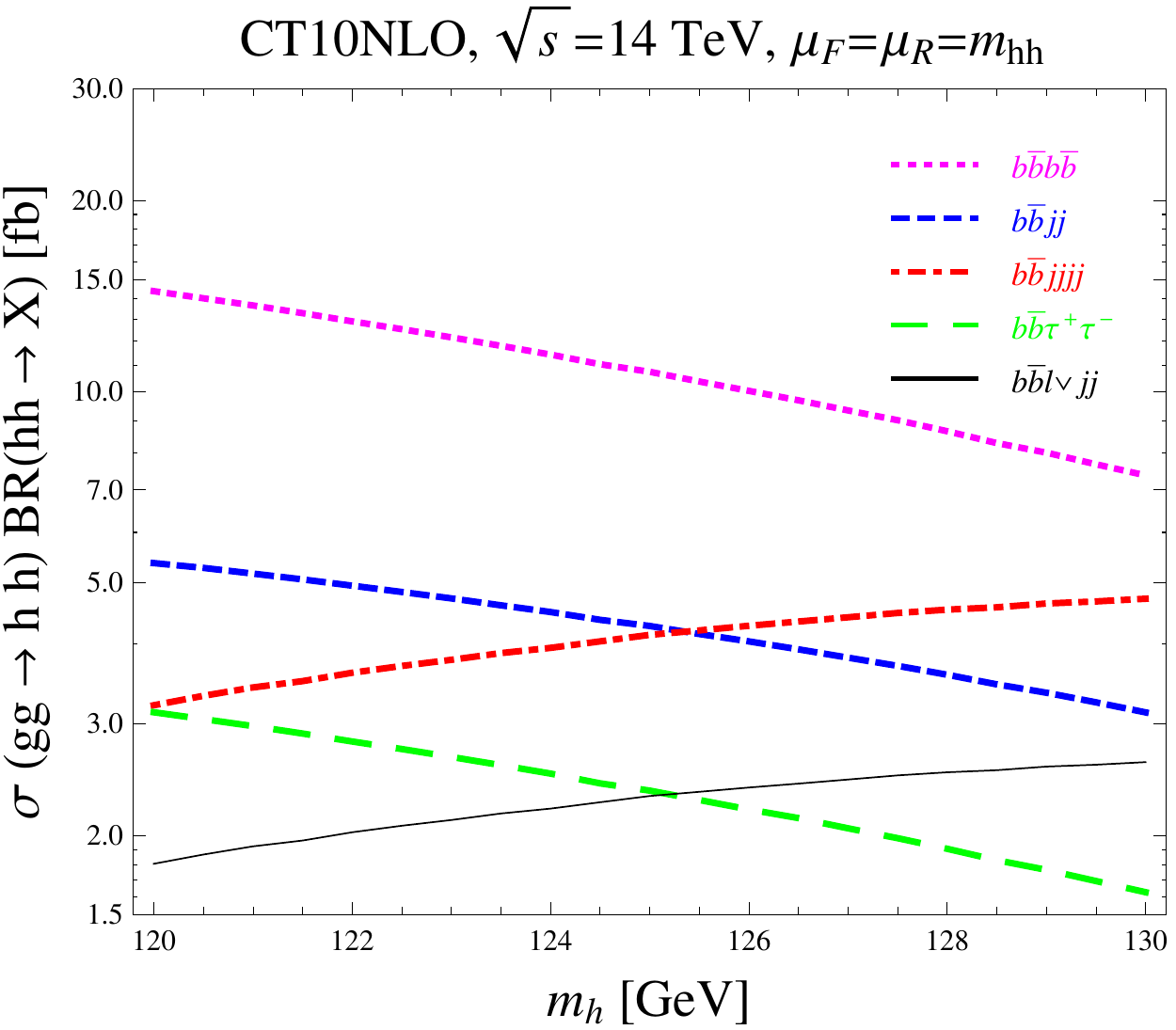}}
\end{minipage}
\caption{Feynman diagrams for $HH$ production at leading order ($f$ is a generic fermion):  (a) triangle and (b) box topologies. (c) Rates (cross section times branching ratios) for $HH$ at the 14 TeV LHC. }
\label{fig:dia}
\end{figure}

As in single Higgs production, the gluon-fusion channel is the dominant production mode ($ \ge 90  \%$)~\cite{Djouadi:1999rca} 
and it also has a large K-factor (NLO over LO cross section). For a long time the cross section was only known up to NLO in the large top mass limit~\cite{Dawson:1998py}
\footnote{ A recent study~\cite{Grigo:2013rya} suggests the error to be $\mathcal{O}(10 \%)$ .}
. 
Recently the NNLL resummation (matched to NLO)~\cite{Shao:2013bz} and the soft-virtual NNLO~\cite{deFlorian:2013uza}  corrections were computed. 
The gluon fusion cross section has two diagrams, which are shown in Fig.~\ref{fig:dia}. The left diagram corresponds to single Higgs boson production, with the $s$-channel $H$ going into $HH$, where the trilinear coupling appears. The other diagram is simply a fermionic box.

We consider a toy model where $\lambda = \lambda_{HHH} / \lambda_{HHH}^{SM}$ and $y_f = g_{H f \bar{f}} / g_{H f \bar{f}}^{\rm}$ can be different from unity~\footnote{In a generic extension of the SM, new particles running in the loops, and also new couplings like $ffHH$ could be present. Here, for simplicity, we take these effects to be subdominant with respect to changes in $y_t$ and $\lambda$. }. Using HPAIR~\cite{HPAIR} we fit  the cross section $\sigma_{HH}$, which reads (using MSTW 2008~\cite{Martin:2009iq})
\begin{equation}\label{eq:xsNLO}
\sigma^{NLO}_{HH} [{\rm fb}] = 9.66 \lambda^2 y_t^2 - 49.9 \lambda y_t + 70.1 y_t^4 + {\cal O} (y_b y_t^2) \, ,
\end{equation}
where we have omitted the bottom loops terms, which have a negligible impact (0.2  \%).They are included in our numerical analysis.
From Eq.~\ref{eq:xsNLO} one sees that the triangle diagram is subdominant (due to the off-shellness of the s-channel $H$), and that the interference with the box diagram is destructive and large. The box-amplitude squared (which has the largest coefficient) scales as $y_t^4$, and thus $\sigma_{HH}$ is very sensitive to $y_t$. As a function of $\lambda$, $\sigma_{HH}$ is a parabola, with a minimum around $\lambda \sim 2.4$, and  symmetric with respect to this value.

We turn now our attention to the decay modes. Given that ($BR(b \bar{b}) \sim 60 \%$), all the relevant channels will include one Higgs boson decaying into $b \bar{b}$. The rates (cross section times branching ratios) for the five most important channels are shown in the right panel of Fig.~\ref{fig:dia}.

Since the channels involving only jets suffer from large QCD multijet backgrounds, then $b \bar{b} \tau^+ \tau^-$ and $b \bar{b} l \nu j j$ ($l =  e, \mu$) offer the best prospects, with a rate of about 2.5 fb for $m_H = 125~\rm{GeV}$. For the $b \bar{b} \gamma \gamma$ the rate is 0.087 fb, however the clean $\gamma \gamma$ final state compensates for the low rate. Before 2012 this was supposed to be the best channel to measure the Higgs self-coupling. With 600 fb$^{-1}$, one expects 6 signal (S) and between 11 and 14 background (B) events~\cite{Baur:2003gp}, depending on the exact assumptions for the background (fake rates, luminosities, efficiencies, etc). Further viable channels, like $\tau \tau W W$ or $b \bar{b} Z Z$ have rates which are an order of magnitude smaller than those mentioned before, and thus it is very challenging to use them to measure $\lambda$ (if not impossible).

During 2012, the detection of Higgs pairs was considered using the BDRS substructure techniques~\cite{Butterworth:2008iy},  which were developed to study $VH, H \to b \bar{b}$. In this process, $H$ recoils against the gauge boson $V$, and gets boosted from this recoil, allowing a better signal to background discrimination for the $H \to b \bar{b}$. In Higgs pair production, one has a similar configuration: the Higgs that decays into $b \bar{b}$ recoils against the other Higgs boson. A thorough study~\cite{Dolan:2012rv} found the best channel to be $b \bar{b} \tau^+ \tau^-$, with $S=57$ and $B=119$ for 600 fb$^{-1}$, considering $p_T^{H} > 100$ GeV. By taking a larger $p_T$ cut the $b \bar{b} l \nu j j$ becomes also feasible~\cite{Papaefstathiou:2012qe}, with $S=9$ and $B=6$~\footnote{Here $l= e, \mu$. Including the $\tau$ one multiplies both S and B by $1 + \epsilon_{\tau}/2$, with $\epsilon_{\tau}$ being the $\tau$ efficiency.}.  

Since both  single and double Higgs production are similar processes (gluon-initiated, uncolored final state) one can expect the cross section ratio $C_{HH} = \sigma_{HH} / \sigma_{H}$ to be more stable against QCD corrections~\cite{Djouadi:2012rh}.
We found that the scale uncertainty on $C_{HH}$ is 9  \% (1.5  \%) at NLO (LO) (to be compared to the 20 (30)  \% for $\sigma_{HH}$).
Moreover, common systematic errors (like the luminosity uncertainty), will cancel out in the ratio. 


A common approach to measure a parameter is to fit the shape of some distribution, which has to be sensitive to the parameter one wants to extract. For instance, for the $\bar{b} b \gamma \gamma$ case one could look at the visible mass distribution, $m_{vis}$. The drawback is that shapes are not always well-known, and moreover having only a handful of events to distribute in a few bins, one might lack enough statistics as to perform an accurate measurement. Therefore, our suggestion~\cite{Papaefstathiou:2012qe} is to do a simple counting experiment. Assuming the number of signal and background events follows a Gaussian distribution, we can build the following (measurable) quantity,
\be
C_{HH}^{\rm exp} = \frac{\sigma_{HH}^{b \bar{b} xx}}{2~\sigma_H^{b \bar{b}}~BR(xx)}
\ee
where $\sigma_{HH}^{b \bar{b} xx} = 2 \sigma_{HH} BR(b \bar{b}) BR(xx)$ and $\sigma_H^{b \bar{b}} =  \sigma_H BR(b \bar{b})$ are measured by the experiment, with $BR(xx)$ being the Higgs branching ratio into the $xx$ final state.
At this point one needs to make some assumptions about the uncertainties. Following Ref.~\cite{cmseuropean} and trying to be conservative, we take a 20  \% uncertainty for the Yukawa top, and for $BR(\tau \tau, WW, \gamma \gamma)$ = (12,12,16)  \%. We further assume no improvement beyond 300/fb (what would happen if systematic errors dominate) and combine all errors in quadrature. In order to put all the channels in the same footing, we rescale the results of Ref.~\cite{Dolan:2012rv} by considering $\epsilon_{\tau} = 0.7$ and $\sigma_{HH} = 31.76$ fb. The number of signal and background events for each channel, and the resulting relative errors $\delta C_{HH} = \Delta C_{HH} / C_{HH}$ are shown in Table~\ref{tab:deltachh}. The PDF uncertainty (about 5  \%) was computed following Ref.~\cite{Watt:2011kp}.

\begin{table}[!htb]
\caption{Number of expected signal and background events and the fractional uncertainties on the ratio of double-to-single
  Higgs boson production cross sections $\delta C_{HH} = \Delta C_{HH} / C_{HH}$, for the different channels and
  the two investigated LHC luminosities, 600~fb$^{-1}$ and
  3000~fb$^{-1}$, using $M_H = 125$~GeV. }
\label{tab:deltachh}
\
\begin{center}
\begin{tabular}{|c|c|c|c|c|} \hline
Process & S (600~fb$^{-1}$) & B (600~fb$^{-1}$) & $\delta C_{HH}$ (600~fb$^{-1}$) &
$\delta C_{HH}$  (3000~fb$^{-1}$)  \\ \hline
$b\bar{b} \tau^+ \tau^-$ & 50 & 104  & 0.400 & 0.279 \\ \hline
$b\bar{b} W^+ W^-$ & 12 & 8  & 0.513 & 0.314\\ \hline
$b\bar{b} \gamma \gamma$ & 6 & 12.5 & 0.964 & 0.490\\ \hline
\end{tabular}
\end{center}
\end{table}

\begin{figure}
\begin{minipage}{0.33\linewidth}
\centerline{\includegraphics[width=0.9\linewidth]{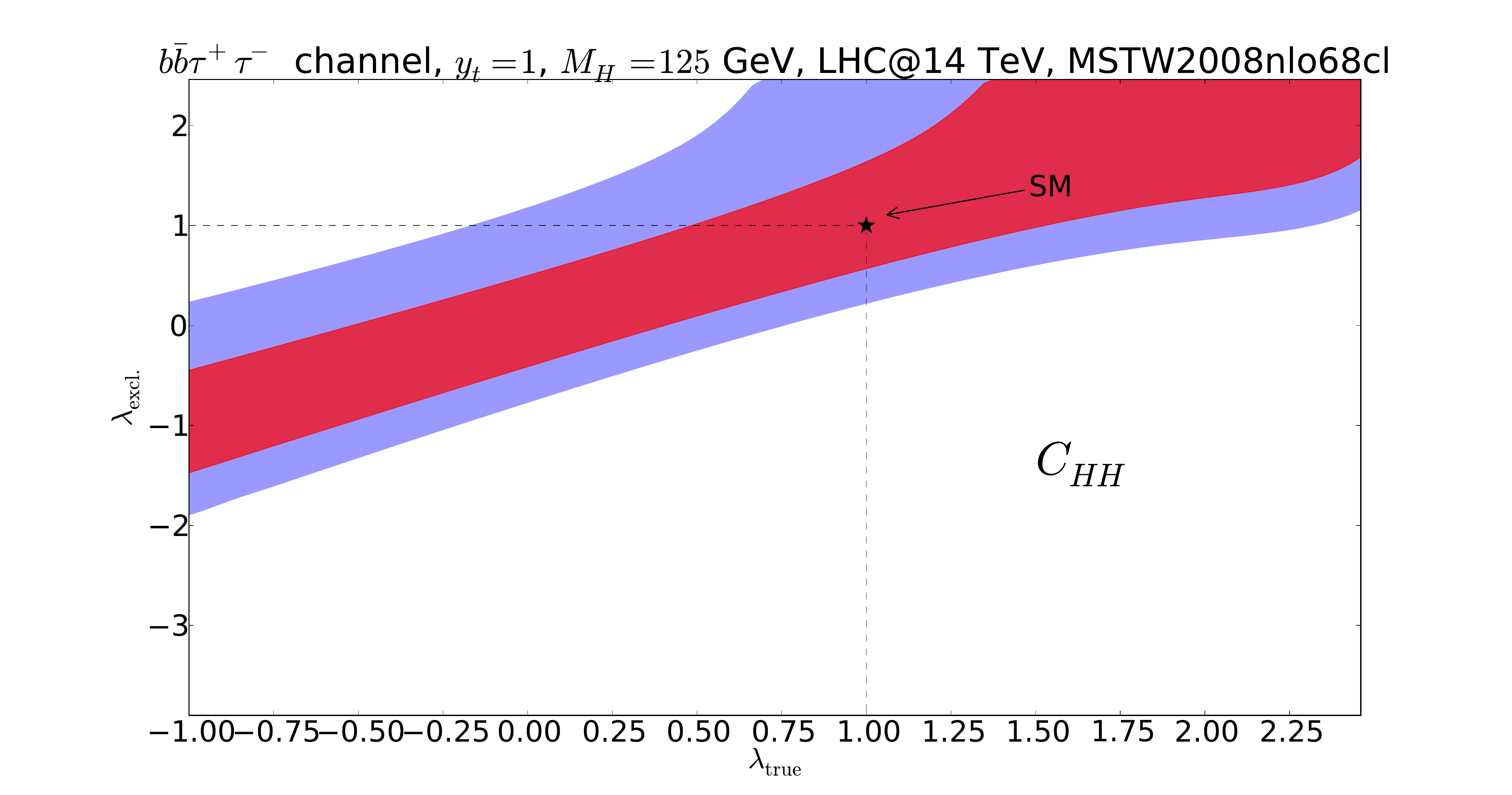}}
\end{minipage}
\begin{minipage}{0.32\linewidth}
\centerline{\includegraphics[width=0.9\linewidth]{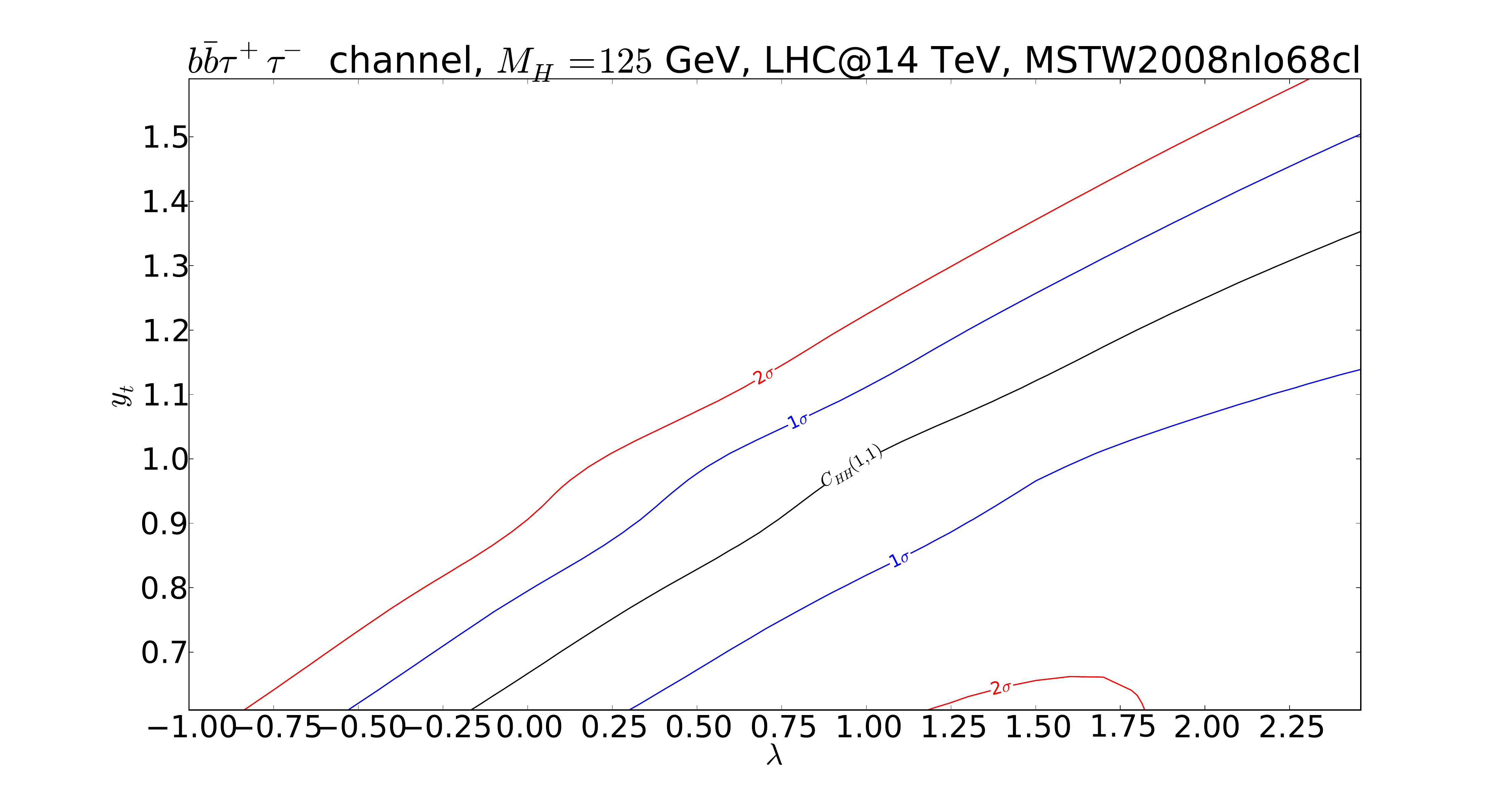}}
\end{minipage}
\hfill
\begin{minipage}{0.32\linewidth}
\centerline{\includegraphics[width=0.9\linewidth]{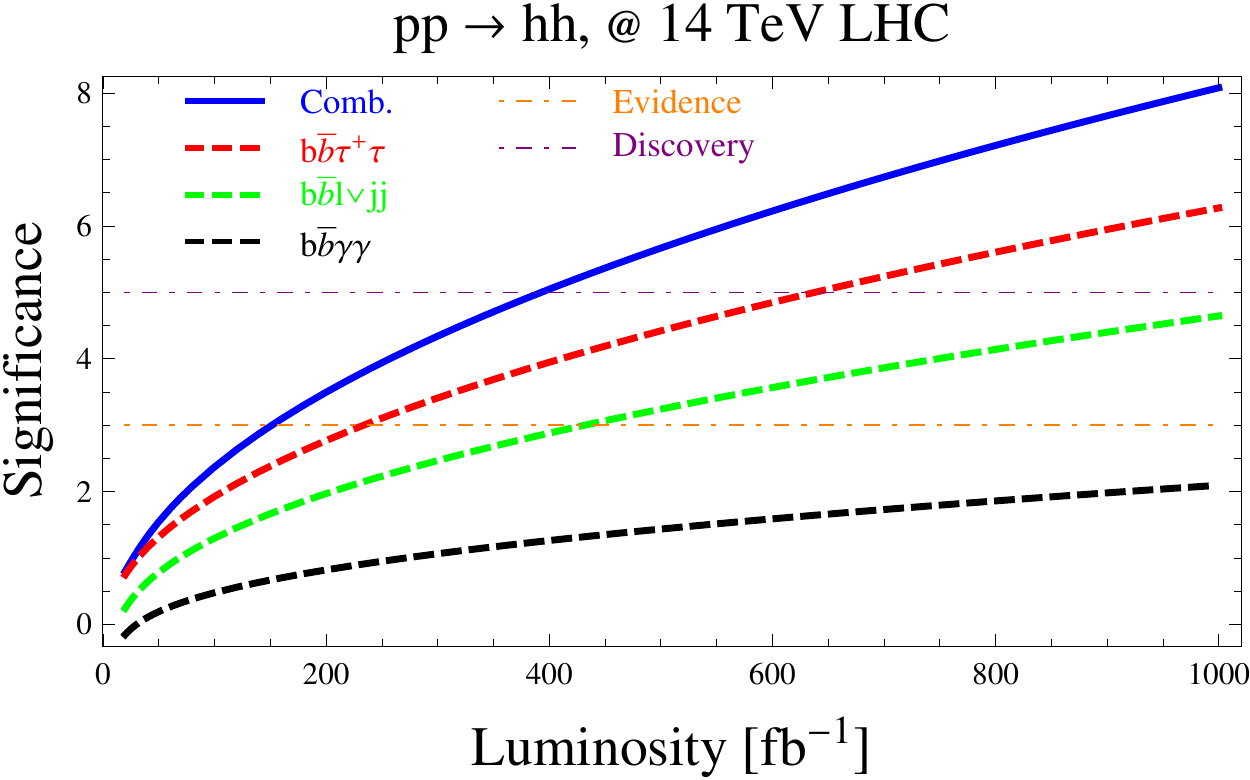}}
\end{minipage}
  \caption{The expected exclusions intervals at one and two standard deviations with 600 ~fb$^{-1}$ for (a) $\lambda_{\rm true} \ne 1$ and (b) for the SM in the $\lambda- y_t$ plane. (c) Significance for each channel and for their combination.}
\label{fig:2}
\end{figure}

Using $\delta C_{hh}$, one can, given an assumption for the true value of $\lambda$ ($\lambda_{\rm true}$) compute the expected confidence interval at a desired confidence level. We show in Table~\ref{tab:excl} the
the 68  \% and 95  \% C.L intervals for the SM case ($\lambda_{\rm true} = 1 = y_{t, {\rm true}}$). A naive combination of all channels, will give a $^{+30}_{-20}$  \% uncertainty, for the 14 TeV LHC with 3000/fb. 
\begin{table}[!htb]
\caption{The expected limits at 1$\sigma$ and 2$\sigma$ confidence
  levels, for $\lambda_\mathrm{true} = 1= y_{t,\mathrm{true}}$  for 600~fb$^{-1}$ and 3000~fb$^{-1}$.}
\label{tab:excl}
\begin{center}
\begin{tabular}{|l|c|c|c|c|} \hline
Process & 600~fb$^{-1}$ (2$\sigma$) & 600~fb$^{-1}$ (1$\sigma$) &
3000~fb$^{-1}$ (2$\sigma$) &  3000~fb$^{-1}$ (1$\sigma$) \\ \hline
$b\bar{b} \tau^+ \tau^-$ &  (0.22, 4.70) & (0.57, 1.64) & (0.42, 2.13) & (0.69, 1.40)\\ \hline
$b\bar{b} W^+ W^-$ & (0.04, 4.88) & (0.46, 1.95)  & (0.36,
4.56)  & (0.65, 1.46)  \\ \hline
$b\bar{b} \gamma \gamma$ &  (-0.56, 5.48) & (0.09, 4.83) & (0.08,
4.84) & (0.48, 1.87)  \\ \hline
\end{tabular}
\end{center}
\end{table}
We consider now the case $\lambda_{\rm true} \ne 1$. The exclusion intervals thus obtained are shown in the left panel of Fig.~\ref{fig:2}. For $\lambda > 1$ the intervals become wider, due to the reduced cross section.

We can also look at the constraints in the $\lambda - y_t$ plane, for the SM case ($y_{t, \rm true} = \lambda = 1$), which are shown in the middle panel of Fig.~\ref{fig:2}. From here we can see the interplay between both couplings. If we assume $\lambda_{\rm true} = 1$ and a 15  \% error in $y_t$~\cite{Peskin:2012we}, the $1 \sigma$ interval is $\lambda \in [0.2-1.1]$, while for $y_t = 1.15$ one has $\lambda \in [1.1-2.5]$, and the $1 \sigma$ intervals for this channel do barely overlap. This shows that any serious attempt to measure the Higgs self-trilinear coupling must also require a very good precision on $y_t$~\footnote{This measurement can also be done with the aid of jet-substructure~\cite{Plehn:2009rk}.}.

It is interesting to know, aside of the measurement of $\lambda$, how much luminosity will the 14 TeV LHC need to discover Higgs pair production. In the right panel of Fig.~\ref{fig:2} we show the significance as a function of the luminosity for the three channels used in our study, and show their combination (combining the significances in quadrature). We see that while the $b \bar{b} \tau \tau$ channel can lead to \emph{evidence} ($3 \sigma$) with 240 fb$^{-1}$ and \emph{discovery} $5 \sigma$ with about 640 fb$^{-1}$, these numbers get reduced to 150 fb$^{-1}$ and 390 fb$^{-1}$ when using the combination~
\footnote{The $b \bar{b} \gamma \gamma$ was recently revisited~\cite{Baglio:2012np}, obtaining $6.5 \sigma$ with 3000 fb$^{-1}$. We do not consider it here, since they neglect the background from jets and electrons misidentified as photons (\emph{fake photons}), which is the dominant one~\cite{Dolan:2012rv}.}.

To summarize, we have studied the extraction of the Higgs self-coupling at the 14 TeV LHC, by exploiting the double-to-single cross section ratio. We have shown that the ratio has a reduced theoretical (scale) uncertainty, and hence is more suitable to derive a limit on $\lambda$ than the pair production cross section itself ($20-30  \%$ theory uncertainty). Including theoretical and experimental uncertainties we have derived a limit on $\lambda$ using the $b \bar{b} \gamma \gamma$,$b \bar{b} WW$ and $b \bar{b} \tau \tau$ channels. A naive combination of the three channels yields a  $^{+30}_{- 20}$  \% accuracy with 3000 fb$^{-1}$. We have also stressed the interplay between $y_t$ and $\lambda$, and the need to have an accurate determination of $y_t$ in order to seriously consider tackling the determination of $\lambda$. The last word will come from ATLAS and CMS, who are the only ones who can perform a full-fledged analysis, including detector simulation, underlying event, pile up, efficiencies, etc.

 \section{Acknowledgments}
J.Z would like to thank the organizers for the invitation to speak at this conference, and for the lively atmosphere during the event.


\bibliography{Zurita_blois.bib}

\end{document}